\documentclass[prd,twocolumn,showpacs,floatfix,amsmath,nofootinbib,floatfix]{revtex4}
\usepackage[utf8x]{inputenc}
\usepackage{dcolumn,booktabs,bm,multirow}
\usepackage{longtable,lscape}
\usepackage{txfonts}
\usepackage{overpic} 
\usepackage{amssymb}
\usepackage{array}
\usepackage{indentfirst}
\usepackage{feynmf}   
\usepackage{slashed}  
\usepackage{cases}
\usepackage{color}
\usepackage{multirow}
\usepackage{setspace}
\usepackage{graphicx}
\usepackage[colorlinks,
            citecolor=green,
            anchorcolor=red,
            menucolor=red,
            linkcolor=red,
            filecolor=red,
            runcolor=red,
            urlcolor=blue,
            frenchlinks=red]{hyperref}
\usepackage{epstopdf}

\begin{document}
\title{$D$ wave bottomonia production from $Z_b^{(\prime)}$ decay}
\author{Qi Wu$^{1}$}\email{wuq@seu.edu.cn}
\author{Dian-Yong Chen$^{1}$\footnote{Corresponding author}}\email{chendy@seu.edu.cn}
\author{Takayuki Matsuki $^{2}$}\email{matsuki@tokyo-kasei.ac.jp}
\affiliation{
$^1$ School of Physics, Southeast University, Nanjing 210094, People's Republic of China\\
$^2$Tokyo Kasei University, 1-18-1 Kaga, Itabashi, Tokyo 173-8602, Japan
}
\date{\today}
\begin{abstract}
In the present work, we investigate the dipion transitions between $\Upsilon(5S)$ and $\Upsilon_J(1D)$ with $J=1, 2, 3$. Our analysis indicates that the dominant sources of the anomalously large widths of $\Upsilon(5S)\to \Upsilon_J(1D) \pi^+ \pi^-$ should be $Z_b^{(\prime)}$, i.e., the dipion transitions occur via the cascade decays $\Upsilon(5S)\to Z_b^{(\prime)\pm}  \pi^\mp \to \Upsilon_J(1D) \pi^+ \pi^-$.  With the assumption that all the short ranged dynamics could be absorbed in a single cutoff with a model parameter $\alpha$, the present estimations indicate that in a reasonable parameter range the measured branching ratios of $\Upsilon(5S) \to \Upsilon_J(1D) \pi^+ \pi^-$ can be reproduced in magnitude, which further proves that the decays via $Z_b^{(\prime)}$ dominate the dipion transitions of $\Upsilon(5S)$ to $\Upsilon_J(1D)$. Moreover, we also predict the ratios of the branching fractions of $Z_b^{(\prime)} \to \Upsilon_J(1D) \pi$, which in our calculations are largely independent of the parameter $\alpha$ and could be tested by further experiments in Belle II.
\end{abstract}
\maketitle

\section{Introduction}
\label{sec1}
A large data sample of $e^+e^-$ collision at the energy of the $\Upsilon(5S)$ resonance has been collected by the Belle collaboration. Based on this large data sample,  the collaboration has reported a series of precise analyses of $\Upsilon(5S)$ decay and has observed some unexpected phenomena, such as the anomalous transition widths between $\Upsilon(5S)$ and lower bottomonia~\cite{Abe:2007tk, Tamponi:2018cuf,Adachi:2011ji, Yin:2018ojs, He:2014sqj}, and the bottomonium-like exotic states $Z_b^{(\prime)}$~\cite{Belle:2011aa, Garmash:2015rfd}. $\Upsilon(5S)$ has become one of intriguing sources to investigate lower bottomonia, such as $\chi_{bJ}$, $h_b$, and $\Upsilon(1D)$.

In Ref.~\cite{Adachi:2011ji}, the Belle collaboration reported the first observation of the $h_b(1P)$ and $h_b(2P)$ in the dipion missing mass spectrum. Besides these two $P$ wave states, the signal of $\Upsilon(1D)$ state is also observed, i.e, the transition process $\Upsilon(5S) \to \Upsilon(1D) \pi^+ \pi^-$ was observed. Moreover, the yields from the fits to the dipion missing mass distributions are $(104.9 \pm 5.8 \pm 3.0) \times 10^3$, $(143.7 \pm 8.7 \pm 6.8) \times 10^3$, and $(22.4 \pm 7.8)\times 10^3$ for $\Upsilon(1S)$, $\Upsilon(2S)$, and $\Upsilon(1D)$, respectively ~\cite{Adachi:2011ji}. The dipion transition branching ratios for $\Upsilon(1S)$ and $\Upsilon(2S)$ are measured to be $\mathcal{B}(\Upsilon(5S)\to \Upsilon(1S) \pi^+ \pi^-)= (5.3 \pm 0.6) \times 10^{-3}$ and $\mathcal{B}(\Upsilon(5S)\to \Upsilon(2S) \pi^+ \pi^-)= (7.8 \pm 1.3) \times 10^{-3}$~\cite{Tanabashi:2018oca}, respectively. Then, one can roughly estimate the branching ratio for the dipion transitions of $\Upsilon(5S) \to \Upsilon(1D) \pi^+ \pi^-$ by the yields of $\Upsilon(1S,2S,1D)$ and the branching ratios for $\Upsilon(1S,2S)$, which are $(1.21 \pm 0.47) \times 10^{-3} $ and $(1.13 \pm 0.42) \times 10^{-3}$, respectively. How to understand such large branching ratios becomes an intriguing problem.

 In the heavy quarkonia dipion transition processes, the dipion can be hadronized by  gluon emitted from the heavy (anti-)quark, this process can be well described by QCD multipole expansion (QCDME) approach~\cite{Bhanot:1979af, Peskin:1979va, Voloshin:1978hc, Bhanot:1979vb, Voloshin:1980zf, Yan:1980uh, Kuang:1990kd}. Such a mechanism plays essential role in the light meson transitions between lower heavy quarkonia . However, the measured widths of the hidden bottom decays of $\Upsilon(5S)$ are much larger than the QCDME expectations~\cite{Abe:2007tk, Tamponi:2018cuf,Adachi:2011ji, Yin:2018ojs, He:2014sqj}.  Since $\Upsilon(5S)$ dominantly decays into a pair of bottom mesons~\cite{Tanabashi:2018oca}, the bottom meson pair can transit into a bottomonia and a light meson by exchanging a proper bottom meson. Such kind of bottom meson loops mechanism plays a crucial role in understanding hidden bottom decays of $\Upsilon(5S)$~\cite{Meng:2007tk,Meng:2008dd, Meng:2008bq, Chen:2011qx, Chen:2014ccr, Wang:2016qmz, Wu:2018xaa}. In Ref.~\cite{Meng:2007tk}, the authors introduced the bottom meson loops to interpret the anomalous widths of $\Upsilon(5S) \to \Upsilon(nS) \pi^+ \pi^-,\ n=\{1,2,3\}$. The decay widths of $\Upsilon(5S) \to \Upsilon(nS) \pi^+ \pi^-,\ n=\{1,2,3\}$ can be well reproduced with meson loop contributions~\cite{Meng:2007tk}, where the dipion is connected to the bottom meson loops via scalar mesons, such as $\sigma$ and $f_0(980)$. Different from the dipion transitions between $\Upsilon(5S)$ and $\Upsilon(nS),\ n=\{1,2,3\}$, the meson loop contributions via a scalar light meson are strongly suppressed in the heavy quark limit since the angular momentum of $\Upsilon(5S)$ and $\Upsilon(1D)$ are $L=0$ and $L=2$, respectively. Thus, the dominant contributions of the meson loops should come from tensor mesons, such as $f_2(1270)$. However, the mass difference between $\Upsilon(5S)$ and $\Upsilon(1D)$ is about 700 MeV, which is much smaller than the mass of $f_2(1270)$, then the contribution from bottom meson loops with $f_2(1270)$ should be strongly suppressed. The above analysis indicates that the sources of anomalously large widths of $\Upsilon(5S) \to \Upsilon(nS) \pi^+ \pi^-$ and $\Upsilon(5S) \to \Upsilon(1D) \pi^+ \pi^-$ should be different.

Although the meson loop contributions can interpret the widths of $\Upsilon(5S) \to \Upsilon(nS) \pi^+ \pi^-$, the dipion invariant mass spectrum and the helicity angle distributions can not be reproduced simultaneously \cite{Chen:2011qx}, especially for $\Upsilon(5S) \to \Upsilon(2S) \pi^+ \pi^-$ process, which is named the $\Upsilon(2S)$ anomaly~\cite{Chen:2011qx, Chen:2011zv}. Besides the anomalous widths, another important phenomenon observed in $\Upsilon(5S)$ decay is the bottomonium-like states $Z_b^{(\prime)}$~\cite{Belle:2011aa, Garmash:2015rfd}. The particular properties of $Z_b^{(\prime)}$ have inspired theorists great interests. Since  $Z_b^{(\prime)}$ was observed in the $\Upsilon(nS) \pi, \ (n=1,2,3)$ and $h_b(mP) \pi, \ (m=1,2)$ invariant mass distributions, thus the isospins of these two states are one. The most possible quark components are $q\bar{q} b\bar{b}$, which indicates these two states could be good candidates of tetraquark states~\cite{Guo:2011gu,Ali:2011ug,Cui:2011fj,Wang:2013zra}. Moreover, $Z_b$ and $Z_b^{\prime}$ are very close to the thresholds of $B^\ast \bar{B}$ and $B^\ast \bar{B}^\ast$, respectively, then authors in Refs.~\cite{Sun:2011uh, Bondar:2011ev, Li:2012wf,Yang:2011rp,Zhang:2011jja,Wang:2013daa,Wang:2014gwa,Dong:2012hc,Ohkoda:2013cea,Liu:2014eka,Li:2012as,Cleven:2013sq,Li:2012uc,Mehen:2011yh, Li:2014pfa, Valderrama:2012jv,Voloshin:2013ez,He:2014nya,Dias:2014pva, Chen:2015jgl, Chen:2016mjn,Abreu:2016xlr,Baru:2017gwo,Wang:2018jlv, Wang:2018pwi, Goerke:2017svb, Ohkoda:2012rj} considered these two states as molecular states composed of $B^\ast \bar{B}$ and $B^\ast \bar{B}^\ast$, respectively. Besides these resonance scenarios, in Refs.~\cite{Chen:2011pv,Chen:2012yr,Bugg:2011jr}, $Z_b$ and $Z_b^\prime$ were interpreted as some special mechanisms, such as initial single pion emission mechanism~\cite{Chen:2011pv,Chen:2012yr} and cusp effects~\cite{Bugg:2011jr}.

It should be stressed that $Z_b$ and $Z_b^\prime$  play important roles in understanding the $\Upsilon(2S)$ anomaly \cite{Chen:2011qx}, which indicates that in the dipion transitions between $\Upsilon(5S)$ and $\Upsilon(nS),\ \{n=1,2,3\}$, both the meson loop contributions and $Z_b^{(\prime)}$ are important in interpreting the anomalous widths. However, as indicated above, the meson loop contributions to the dipion transitions between $\Upsilon(5S)$ and $\Upsilon(1D)$ should be suppressed. Hence, the most possible sources of the anomalous width of $\Upsilon(5S) \to \Upsilon(1D) \pi^+ \pi^-$ are $Z_b^{(\prime)}$, i.e., the transition occurs via the cascade decays $\Upsilon (5S) \to Z_b^{(\prime) \pm}  \pi^\mp \to \Upsilon(1D) \pi^\pm \pi^\mp$.

In Ref.~\cite{Wu:2018xaa}, the decay processes $\Upsilon(5S) \to Z_b^{(\prime)} \pi$ have been investigated, and the branching ratios of $\Upsilon(5S) \to Z_b^{(\prime)} \pi$ are estimated to be of order $1\%$, which is consistent with the experimental measurements by the Belle collaboration. In Ref.~\cite{Xiao:2017uve}, the decays of $Z_b^{(\prime)}$ were estimated via meson loop mechanism by using the fact that $Z_b^{(\prime)}$ dominantly decay into a pair of bottom mesons, which is similar to the higher bottomonia. In the present work, we estimate the meson loop contributions to the widths of $Z_b^{(\prime)} \to \Upsilon(1D) \pi$, and together with the estimation in Ref.~\cite{Wu:2018xaa}, we will further check the role of $Z_b^{(\prime)}$ played in $\Upsilon(5S) \to \Upsilon(1D) \pi^+ \pi^- $ process. Moreover, ever since the observation of $\Upsilon(1D)$, the masses of spin triplets have not been well measured because the mass splitting of the spin triplet is small. Thus, in the measurement of the dipion missing mass spectrum, only one peak around 10160 MeV was observed, which should contains $\Upsilon_1(1D)$, $\Upsilon_2(1D)$, and $\Upsilon_3(1D)$. From the present estimation, we can find dominant components in the observed structure.

This work is organized as follows. After introduction, we present our estimates of the branching ratios of $Z_b^{(\prime)}  \to \Upsilon(1D) \pi $ in an effective Lagrangian approach. The numerical results and discussions are presented in Section \ref{Sec:Num} and the last section is devoted to a short summary.

\begin{figure}[t]
\begin{tabular}{ccc}
 \includegraphics[width=2.7cm]{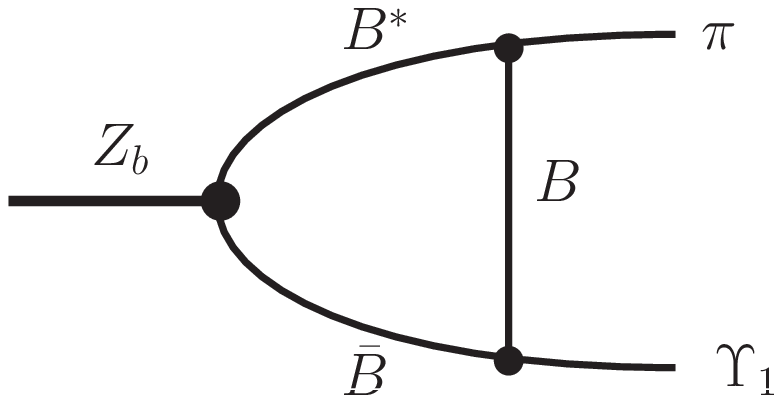}&
 \includegraphics[width=2.7cm]{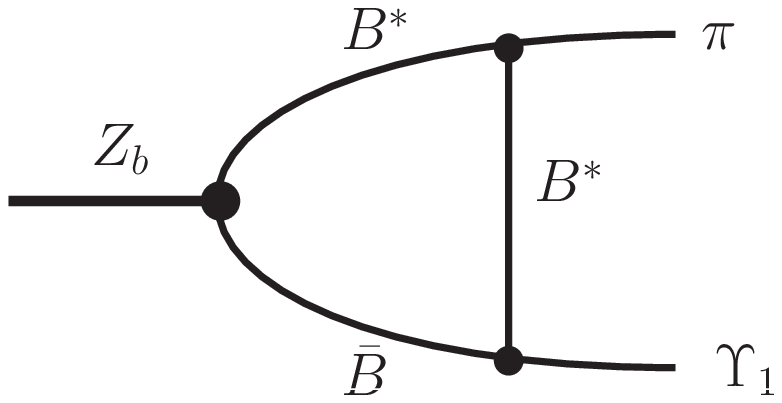}&
 \includegraphics[width=2.7cm]{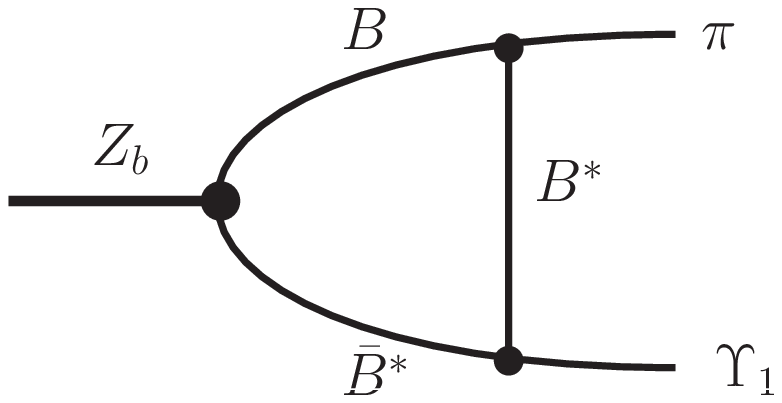}\\
 \\
 $(a)$ & $(b)$ &$(c)$\\
 \\
 \includegraphics[width=2.7cm]{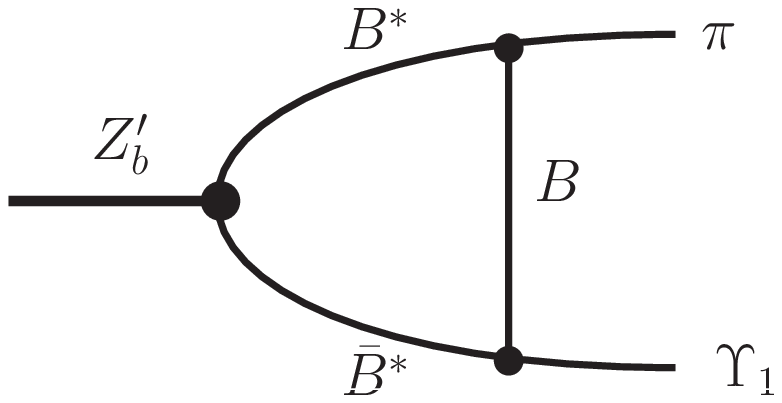}&
 \includegraphics[width=2.7cm]{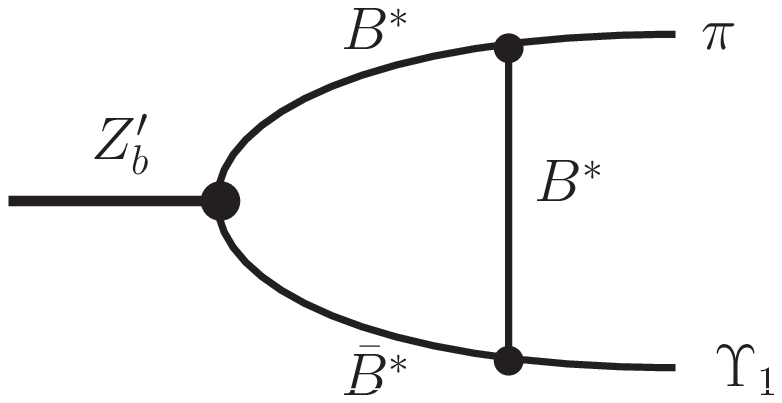}&
 \includegraphics[width=2.7cm]{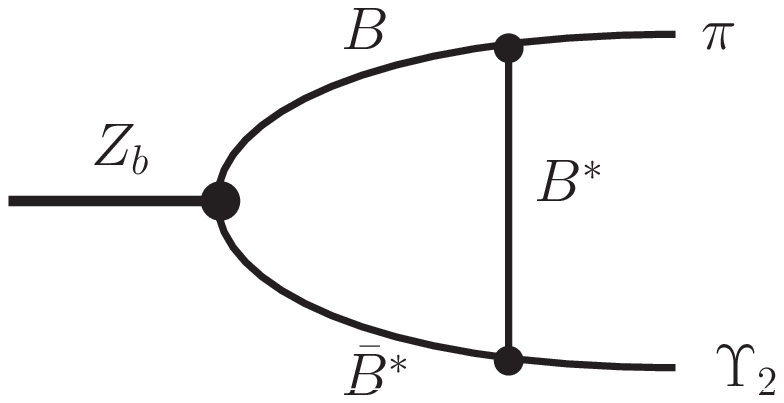}\\
 \\
 $(d)$ & $(e)$& $(f)$\\
 \\
 \includegraphics[width=2.7cm]{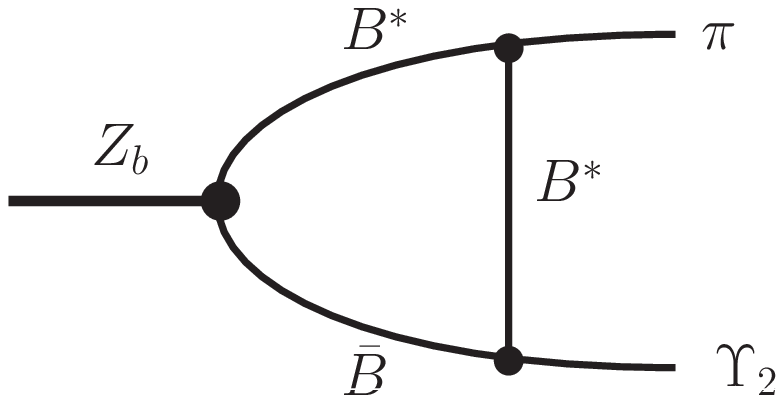}&
 \includegraphics[width=2.7cm]{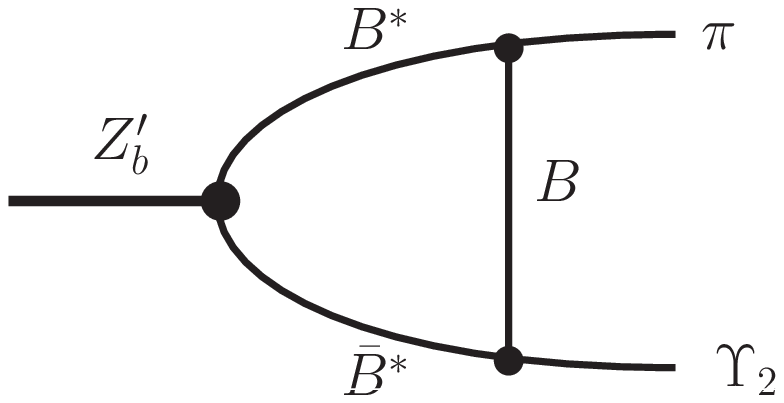}&
 \includegraphics[width=2.7cm]{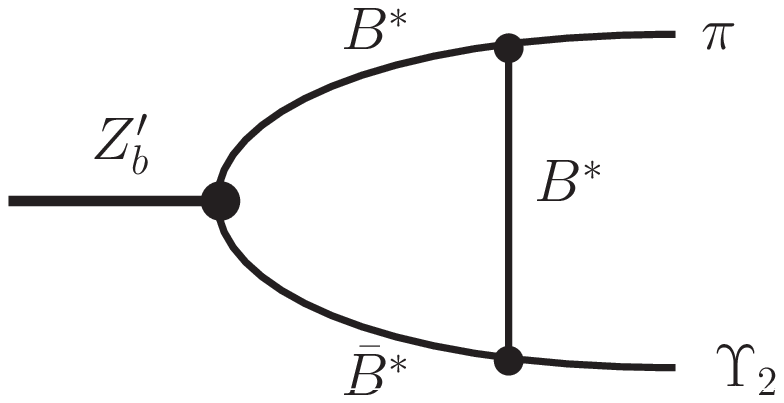}\\
 $(g)$ & $(h)$& $(i)$\\
 \\
 \includegraphics[width=2.7cm]{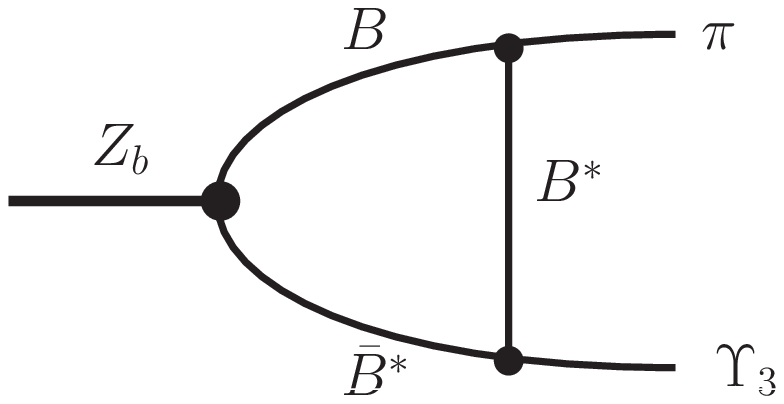}&
 \includegraphics[width=2.7cm]{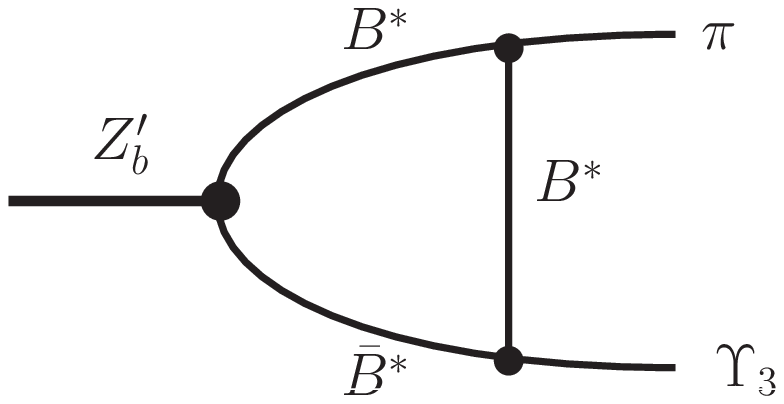}&\\
 \\
 $(j)$ & $(k)$&
 \end{tabular}
  \caption{Diagrams contributing to $Z^{(\prime)}_b\to \Upsilon_J(1D) \pi$.}\label{Fig:Feyn}
\end{figure}

\section{Meson loop contributions to $Z^{(\prime)}_b\rightarrow \Upsilon_J(1D)\pi$   }
\label{sec:For}

The experimental measurements indicate that $Z_b$ and $Z_b^\prime$ dominantly decay into a pair of bottom mesons, in particular, the branching ratios of the open bottom channels are measured to be of $\mathcal{B}(Z_b \to B^\ast \bar{B} +c.c.)=(85.6^{+1.5+1.5}_{-2.0-2.1})\%$ and $\mathcal{B}(Z_b^\prime \to B^\ast \bar{B}^\ast)= (73.7^{+3.4+2.7}_{-4.4-3.5})\% $, respectively~\cite{Garmash:2015rfd}. Similar to the case of higher bottomonia, the bottom meson loops are also expected to play dominant roles in the hidden bottom decays of $Z_b^{(\prime)}$~\cite{Xiao:2017uve}. In Fig.~\ref{Fig:Feyn}, we present all the possible diagrams contributing to $Z_b^{(\prime)} \to \Upsilon_J(1D) \pi$, where diagrams (a)-(e) correspond to $Z_b^{(\prime)} \to \Upsilon_1(1D) \pi$, while diagrams (f)-(i) and (j)-(k) correspond to $Z_b^{(\prime)} \to \Upsilon_2(1D) \pi$ and $Z_b^{(\prime)} \to \Upsilon_3(1D) \pi$, respectively.

All the diagrams in Fig.~\ref{Fig:Feyn} can be estimated at hadronic level in an effective Lagrangian approach. As for the interactions between $Z_b^{(\prime)}$ and bottom meson pair, we can simply consider the $S$ wave coupling and the corresponding effective Lagrangians can be expressed as,
\begin{eqnarray}
\mathcal{L}=g_{Z_b B^\ast B}Z^\mu_b \left(B\bar{B}^\ast_\mu+B^\ast_\mu \bar{B}\right)+ig_{Z^\prime_b B^\ast B^\ast}\epsilon^{\mu\nu\alpha\beta}\partial_\mu Z^\prime_{b\nu}B^\ast_\alpha\bar{B}^\ast_\beta.
\end{eqnarray}

In the heavy quark limit, the wave function of a heavy-light meson has to be independent on the flavor and spin of the heavy quark. Therefore, the heavy-light mesons can be characterized by the light degrees of freedom, i.e., $s_\ell =s_q +\ell$, where $s_q$ and $\ell$ are the spin and orbital angular momentum of a light quark, respectively. In this case, the heavy-light mesons with the same light quark freedom can be degenerate. For example, the $S$-wave doublets $(B,B^\ast)$ and $(\bar{B}, \bar{B}^\ast)$ can be expressed in the matrix form, which are \cite{Casalbuoni:1996pg,Kaymakcalan:1983qq,Oh:2000qr,Falk:1992cx},
\begin{eqnarray}
H_1&=&\frac{1+v\!\!\!\slash}{2}\Big[B^{\ast\mu}\gamma_\mu-B\gamma_5\Big],\nonumber\\	
H_2&=&\Big[\bar{B}^{\ast\mu}\gamma_\mu-\bar{B}\gamma_5\Big]\frac{1+v\!\!\!\slash}{2},
\end{eqnarray}
respectively.
Similarly, the heavy quarkonia with the same orbital angular momentum are also degenerate in the heavy quark limit, and the matrix form of the $D$ wave bottomonia can be expressed as\cite{Casalbuoni:1996pg},
\begin{eqnarray}
\mathcal{J}^{\mu\lambda}&=&\frac{1+v\!\!\!\slash}{2}\bigg[\Upsilon^{\mu\alpha\lambda}_3 \gamma_\alpha+\frac{1}{\sqrt{6}}\left(\epsilon^{\mu\alpha\beta\rho}v_\alpha \gamma_\beta \Upsilon^\lambda_{2\rho}+\epsilon^{\lambda\alpha\beta\rho}v_\alpha \gamma_\beta \Upsilon^\mu_{2\rho}\right)\nonumber\\
&&+\frac{\sqrt{15}}{10}\left[\Big(\gamma^\mu-v^\mu \Big)\Upsilon^\lambda_1+\left(\gamma^\lambda-v^\lambda\right)\Upsilon^\mu_1\right]\nonumber\\
&&-\frac{1}{\sqrt{15}} \left(g^{\mu\lambda}-v^\mu v^\lambda \right)\gamma_\alpha \Upsilon^\alpha_1+\eta^{\mu\lambda}_{b2}\gamma_5\bigg]\frac{1-v\!\!\!\slash}{2}.
\end{eqnarray}
With the above matrix form of the degenerate bottom mesons and bottomonia, one can construct the effective interaction between $D$-wave bottomonia and $S$-wave bottom meson pair, which is~\cite{Casalbuoni:1996pg,Wang:2016qmz}
\begin{eqnarray}
\mathcal{L}=ig_2 \mathrm{Tr}\left[\mathcal{J}^{\mu\lambda}\bar{H}_2 {\stackrel{\leftrightarrow}{\partial_\mu}}\gamma_\lambda \bar{H}_1 \right]+\mathrm{H.c.}, \label{Eq:L1}
\end{eqnarray}
where $\bar{H}_{1,2}= \gamma^0 H^\dagger_{1,2} \gamma^0 $. Expanding the above Lagrangian in the matrix form, one can obtain the specific effective interactions involved with the present calculations, which are,
 \begin{eqnarray}
\mathcal{L}&=&g_{\Upsilon_1 BB}\Upsilon^\mu_1 \Big(B^\dagger \partial_\mu B-B\partial_\mu B^\dagger\Big)\nonumber\\
&&+g_{\Upsilon_1 BB^\ast}\epsilon^{\mu\nu\alpha\beta}\Big(B^\dagger{\stackrel{\leftrightarrow}{\partial_\mu}}B^\ast_\beta-B^{\ast\dagger}_\beta{\stackrel{\leftrightarrow}{\partial_\mu}}
B\Big)\partial_\nu \Upsilon_{1\alpha}\nonumber\\
&&+g_{\Upsilon_1 B^\ast B^\ast}\Big[-4\Upsilon^\mu_1 \Big(B^{\ast\nu}\partial_\mu B^{\ast\dagger}_\nu-B^{\ast\dagger}_\nu \partial_\mu B^{\ast\nu}\Big)+\Upsilon^\mu_1 B^{\ast\nu}\partial_\nu B^{\ast\dagger}_\mu\nonumber\\
&&-\Upsilon^\mu_1 B^{\ast\nu\dagger}\partial_\nu B^\ast_\mu\Big]+ig_{\Upsilon_2 B B^\ast}\Upsilon^{\mu\nu}_2\Big(B^\dagger{\stackrel{\leftrightarrow}{\partial_\nu}}B^\ast_\mu-B^{\ast\dagger}_\mu{\stackrel{\leftrightarrow}{\partial_\nu}}B\Big)\nonumber\\
&&+ig_{\Upsilon_2 B^\ast B^\ast}\epsilon_{\alpha\beta\mu\nu}\Big(B^{\ast\nu\dagger}{\stackrel{\leftrightarrow}{\partial^\beta}}B^{\ast}_\lambda-B^{\ast\nu}{\stackrel{\leftrightarrow}
{\partial^\beta}}B^{\ast\dagger}_\lambda\Big)\partial^\mu \Upsilon^{\alpha\lambda}_2\nonumber\\
&&+g_{\Upsilon_3 B^\ast B^\ast}\Upsilon^{\mu\nu\alpha}_3 \Big (B^{\ast\dagger}_\alpha{\stackrel{\leftrightarrow}{\partial_\mu}}B^\ast_\nu+B^{\ast\dagger}_\nu{\stackrel{\leftrightarrow}{\partial_\mu}}B^\ast_\alpha\Big).
\end{eqnarray}

Considering the heavy quark limit and chiral symmetry, one can construct the effective interactions for bottom mesons and pesudoscalar mesons, which are~\cite{Casalbuoni:1996pg,Yan:1992gz,Cheng:1992xi,Falk:1992cx,Wise:1992hn,Burdman:1992gh},
\begin{eqnarray}
\mathcal{L}&=&-ig_{B^\ast B \mathcal{P}}\Big(B^{i\dagger} \partial^\mu \mathcal{P}_{ij}B^{\ast j}_\mu-B^{\ast i\dagger}_\mu \partial^\mu \mathcal{P}_{ij}B^{j}\Big)\nonumber\\
&&+\frac{1}{2}g_{B^\ast B^\ast \mathcal{P}}\epsilon_{\mu\nu\alpha\beta}B^{\ast \mu\dagger}_i \partial^\nu \mathcal{P}_{ij}{\stackrel{\leftrightarrow}{\partial^\alpha}}B^{\ast \beta}_j,
\end{eqnarray}
where $\mathcal{P}$ is the $3\times3$ matrix for the octet pseudoscalar mesons.

With the above effective Lagrangians, we can obtain the amplitudes for $Z_b\to \Upsilon_1 \pi$ corresponding to diagrams (a)-(c), which are
\begin{eqnarray}
\mathcal{M}_{a}&=&i^3 \int\frac{d^4 q}{(2\pi)^4}\Big[g_{Z_b}\epsilon^\mu_{Z_b}\Big]\Big[ig_{B^\ast B\pi}ip^{\nu}_3\Big]\Big[g_{\Upsilon_1 BB}\epsilon^\alpha_{\Upsilon_1}(-i)(p_{2\alpha}-q_{\alpha})\Big]\nonumber\\
&&\frac{-g_{\mu\nu}+p_{1\mu} p_{1\nu} /m^2_1}{p^2_1-m^2_1}\frac{1}{p^2_2-m^2_2}\frac{1}{q^2-m^2_q}\mathcal{F}^2(q^2,m_q^2)\nonumber\\
\mathcal{M}_{b}&=&i^3 \int\frac{d^4 q}{(2\pi)^4}\Big[g_{Z_b}\epsilon^\kappa_{Z_b}\Big]\Big[\frac{1}{2}g_{B^\ast B^\ast \pi}\varepsilon_{\mu\nu\alpha\beta}ip^\nu_3 i(p^\alpha_1+q^\alpha)\Big]\nonumber\\
&&\Big[g_{\Upsilon_1 B^\ast B}\varepsilon_{\tau\xi\eta\phi}(-i)(p^\tau_2-q^\tau) ip^\xi_4 \epsilon^\eta_{\Upsilon_1}\Big]\frac{-g_\kappa^{\mu}+p_{1\kappa} p^{\mu}_1 /m^2_1}{p^2_1-m^2_1}\nonumber\\
&&\frac{1}{p^2_2-m^2_2}\frac{-g^{\beta\phi}+q^{\beta}q^{\phi}/m^2_q}{q^2-m^2_q}\mathcal{F}^2(q^2,m_q^2)\nonumber\\
\mathcal{M}_{c}&=&i^3 \int\frac{d^4 q}{(2\pi)^4}\Big[g_{Z_b}\epsilon^\mu_{Z_b}\Big]\Big[-ig_{B^\ast B \pi}ip_{3\nu}\Big]\Big[g_{\Upsilon_1 B^\ast B^\ast}\epsilon^\alpha_{\Upsilon_1}(-i)\nonumber\\
&&\Big(-4(q_\alpha-p_{2\alpha}) g_{\tau\xi}+q_\tau g_{\alpha\xi}-p_{2\xi} g_{\alpha\tau}\Big)\Big]\frac{1}{p^2_1-m^2_1}\nonumber\\
&&\frac{-g_\mu^{\tau}+p_{2\mu} p^{\tau}_2 /m^2_2}{p^2_2-m^2_2}\frac{-g^{\nu\xi}+q^{\nu}q^{\xi}/m^2_q}{q^2-m^2_q}\mathcal{F}^2(q^2,m_q^2),
\end{eqnarray}
and obtain the amplitudes for $Z_b^\prime \to \Upsilon_1 \pi$ corresponding to diagrams (d)-(e), which are,
\begin{eqnarray}
\mathcal{M}_{d}&=&i^3 \int\frac{d^4 q}{(2\pi)^4}\Big[ig_{Z^\prime_b}\varepsilon_{\mu\nu\alpha\beta}(-i)p^\mu \epsilon^\nu_{Z^\prime_b}\Big]\Big[ig_{B^\ast B\pi}ip_{3\theta}\Big]\Big[-g_{\Upsilon_1 B^\ast B}\nonumber\\
&&\varepsilon_{\rho\tau\sigma\xi}(-i)(p^\rho_2-q^\rho) ip^\tau_4 \epsilon^\sigma_{\Upsilon_1}\Big]\frac{-g^{\alpha\theta}+p^{\alpha}_1 p^{\theta}_1 /m^2_1}{p^2_1-m^2_1}\nonumber\\
&&\frac{-g^{\beta\xi}+p^{\beta}_2 p^{\xi}_2 /m^2_2}{p^2_2-m^2_2}\frac{1}{q^2-m^2_q}\mathcal{F}^2(q^2,m_q^2)\nonumber\\
\mathcal{M}_{e}&=&i^3 \int\frac{d^4 q}{(2\pi)^4}\Big[ig_{Z^\prime_b}\varepsilon_{\mu\nu\alpha\beta}(-i)p^\mu \epsilon^\nu_{Z^\prime_b}\Big]\Big[\frac{1}{2}g_{B^\ast B\ast \pi}\varepsilon_{\theta\phi\kappa\lambda}ip^\phi_3 i\nonumber\\
&&(p^\kappa_1+q^\kappa)\Big]\Big[g_{\Upsilon_1 B^\ast B^\ast}\epsilon^\rho_{\Upsilon_1}(-i)\Big(-4(q_\rho-p_{2\rho}) g_{\sigma\xi}+q_\sigma g_{\rho\xi}\nonumber\\
&&-p_{2\xi} g_{\sigma\rho}\Big)\Big]\frac{-g^{\alpha\theta}+p^{\alpha}_1 p^{\theta}_1 /m^2_1}{p^2_1-m^2_1}\frac{-g^{\beta\sigma}+p^{\beta}_2 p^{\sigma}_2 /m^2_2}{p^2_2-m^2_2}\nonumber\\
&&\frac{-g^{\lambda\xi}+q^{\lambda}q^{\xi}/m^2_q}{q^2-m^2_q}\mathcal{F}^2(q^2,m_q^2).
\end{eqnarray}
In the above amplitudes, we assume that all the short ranged interactions could be fully absorbed into a single form factor as in Refs.~
\cite{Cheng:2004ru, Tornqvist:1993vu, Tornqvist:1993ng, Locher:1993cc, Li:1996yn, Chen:2014ccr, Wang:2016qmz, Chen:2012nva, Chen:2010re},  which is also introduced to depict the internal structures of the exchanged mesons. Moreover, the form factor also plays a role of removing the divergences in the amplitudes. Here, the form factor is in the form ~\cite{Cheng:2004ru, Tornqvist:1993vu, Tornqvist:1993ng, Locher:1993cc, Li:1996yn, Chen:2014ccr, Wang:2016qmz, Chen:2012nva, Chen:2010re},
\begin{eqnarray}
\mathcal{F}(q^2,m_q^2)= \frac{m_q^2-\Lambda^2}{q^2-\Lambda^2},
\end{eqnarray}
where the parameter $\Lambda$ can be reparameterized as $\Lambda=m_q+\alpha\Lambda_{QCD}$ with $\Lambda_{QCD}=220$ MeV and $m_q$ to be the QCD energy scale and the mass of the exchange bottom meson, respectively. In the same way, the amplitudes of $Z_b^{(\prime)}\to \Upsilon_2 \pi$ and $Z_b^{(\prime)}\to \Upsilon_3 \pi$ can be obtained, which are collected in Appendix~\ref{Sec:App-A}.

With the above amplitudes corresponding to diagrams in Fig.~\ref{Fig:Feyn}, we can obtain the total amplitudes for $Z_b^{(\prime)}\to \Upsilon_J(1D) \pi$. Taking $Z_b\to \Upsilon_1(1D) \pi$ as an example, the total amplitude is
\begin{eqnarray}
\mathcal{M}_{\mathrm{tot}}= \mathcal{M}_a+\mathcal{M}_b+\mathcal{M}_c.
\end{eqnarray}
and the corresponding width is,
\begin{eqnarray}
	\Gamma= \frac{1}{3}\frac{1}{32\pi^2} \left|\overline{\mathcal{M}_{\mathrm{tot}}}\right|^2 \frac{|\vec{p_1}|}{m_0^2} d\Omega
\end{eqnarray}
 where the overline indicates the spin sum over the involved states and $\vec{p}_1$ is the momentum of the final state in the initial rest frame.

\section{Numerical results   }
\label{Sec:Num}

The uncertainty of the measured branching ratios of the open bottom decay channels of $Z_b^{(\prime)}$ are very tiny, thus, we do not take this uncertainty into consideration in the following estimations. With the center values of widths of $Z_b^{(\prime)}$, one can get the partial widths of $Z_b^{(\prime)} \to B^{(\ast)} \bar{B}^\ast +c.c.$, which are $\Gamma(Z_b\to B^\ast \bar{B} +c.c.)=15.8$ MeV and $\Gamma(Z_b^\prime \to B^\ast \bar{B}^\ast) =8.48$ MeV, respectively. With these measured widths, one can get the corresponding coupling constants in Eq.~(\ref{Eq:L1}), which are $g_{Z_b B^\ast B}= 13.07$ GeV and $g_{Z_b^\prime B^\ast B^\ast}=0.99$, respectively  \footnote{It should be mentioned that the masses of $Z_b$ and $Z_b^\prime$ are very close to the thresholds of $B^\ast B$ and $B^\ast B^\ast $, respectively. The effective couplings $g_{Z_b^{(\prime)} B^\ast B^{(\ast)}}$ cannot be calculated from the widths without assumptions on the $Z_b^{(\prime)}$ nature in the strict sense as indicated in Ref.~\cite{Guo:2016bjq}. However, we want to compare our numerical results with the experimental measurement to further prove the dominant role of $Z_b^{(\prime)}$ played in the dipion transitions between $\Upsilon(5S)$ and $\Upsilon(1D)$. Thus, we assume that the coupling constants can be directly estimated from the partial widths in the present work.} .

In the heavy quark limit, the coupling constants of $D$-wave bottomonia and bottom meson pair can be related to a unified coupling constant $g_2$ by the following relations,
\begin{eqnarray}
g_{\Upsilon_1 BB}&=&-2g_2 \frac{\sqrt{15}}{3}\sqrt{m_{\Upsilon_1}m_B m_B},\nonumber\\
g_{\Upsilon_1 BB^\ast}&=&-g_2 \frac{\sqrt{15}}{3}\sqrt{m_{B}m_{B^\ast}/m_{\Upsilon_1}},\nonumber\\
g_{\Upsilon_1 B^\ast B^\ast}&=&g_2 \frac{\sqrt{15}}{15}\sqrt{m_{\Upsilon_1}m_{B^\ast} m_{B^\ast}},\nonumber\\
g_{\Upsilon_2 B B^\ast}&=&2g_2 \sqrt{\frac{3}{2}}\sqrt{m_{\Upsilon_2}m_B m_{B^\ast}},\nonumber\\
g_{\Upsilon_2 B^\ast B^\ast}&=&-2g_2 \sqrt{\frac{1}{6}}\sqrt{m_{B^\ast}m_{B^\ast}/m_{\Upsilon_2}},\nonumber\\
g_{\Upsilon_3 B^\ast B^\ast}&=&2g_2 \sqrt{m_{\Upsilon_3}m_{B^\ast} m_{B^\ast}}.
\end{eqnarray}
where the gauge coupling constant $g_2=9.83 \mathrm{GeV}^{-3/2}$, which was determined by the vector meson dominance ansatz~\cite{Wang:2016qmz}.

Considering the heavy quark limit and chiral symmetry, the coupling constants of $g_{B^{(\ast)}B^{(\ast)}\pi}$ can be related to the gauge coupling $g$ by~\cite{Colangelo:2003sa}
\begin{eqnarray}
g_{B^{\ast}B\pi}=\frac{2g}{f_\pi}\sqrt{m_{B^\ast}m_{B}},\qquad   g_{B^{\ast}B^\ast \pi}=\frac{2g}{f_\pi},
\end{eqnarray}
where $f_\pi=132\ \mathrm{MeV}$ is a pion decay constant and $g=0.5$ is determined by using the partial width of $D^\ast\rightarrow D\pi$.

As discussed in the introduction, the mass of $\Upsilon_J(1D)$ has not been measured except for $\Upsilon_2(1D)$, which is $10.164 \mathrm{GeV}$ \cite{Agashe:2014kda}. In the present work, we take the masses of $\Upsilon_1(1D)$ and $\Upsilon_3(1D)$ as those predicted by the potential model in Ref.~\cite{Eichten:1980mw} , which are  $10.153 \mathrm{GeV}$ and \cite{Agashe:2014kda} and $10.174 \mathrm{GeV}$, respectively.

The masses of $Z_b$ and $Z_b^\prime$ are close to the thresholds of $B^\ast {B}$ and $B^\ast \bar{B}^\ast$, respectively. Following the formulas in Ref. \cite{Guo:2017jvc,Guo:2012tg}, the  velocities of bottom mesons involved in the meson loops can be evaluated as,
\begin{eqnarray}
	v=\frac{v_1+v_2}{2}=0.15\sim0.16,
\end{eqnarray}
which is rather small. In this case, the momentum of a bottom meson can be $p^\mu=(p_0,\vec{p})$ with $p_0 \sim m_{B^{\ast}}$ and $\vec{p} \sim m_b v << m_{B^\ast}$, thus the numerator of a vector bottom meson propagator can be simplified as $-g^{\mu\nu} +p^\mu p^\nu/m_{B^\ast}^2 \simeq \delta^{ij}$.

\begin{figure}[t]
  \centering
 \includegraphics[width=8.0cm]{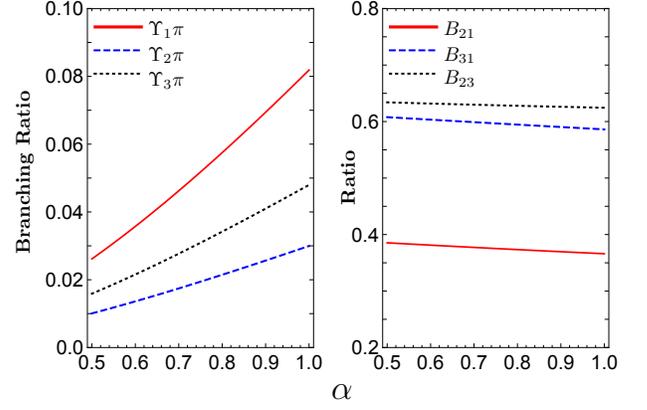}
  \caption{The $\alpha$-dependence of the branching ratios of $Z_b\rightarrow \Upsilon_J(1D)\pi$ (left panel) and of the ratios of the branching ratios for $Z_b\to \Upsilon_J(1D) \pi$ with a different quantum number $J$ (right panel).}\label{Fig:zb}
\end{figure}

\begin{figure}[t]
  \centering
 \includegraphics[width=8cm]{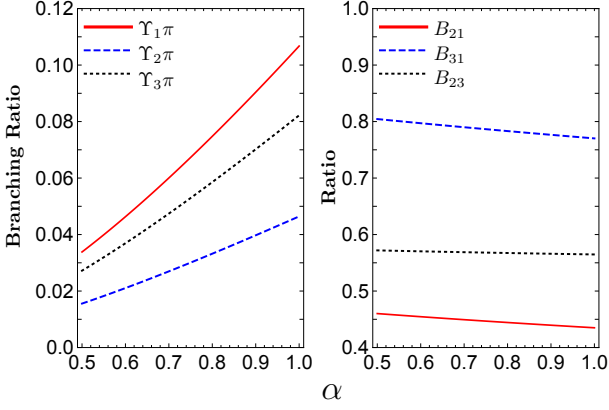}
  \caption{The same as Fig.~\ref{Fig:zbp} but for $Z_b^\prime$.}\label{Fig:zbp}
\end{figure}

In the present estimations, there is only one model parameter $\alpha$ introduced by the form factor. As indicated in Ref.~\cite{Cheng:2004ru} the cutoff $\Lambda$ in the form factor should not be far away from the mass of the exchanged meson. The model parameter $\alpha$ is expected to be of order unity and usually depends on particular processes ~\cite{Cheng:2004ru}. Actually, the model parameter $\alpha$ can not be estimated from the first principle. It is usually determined by comparing the theoretical estimations with the corresponding experimental data, and check the reasonability of the determined parameter range. However, as for the processes involved in the present work, there are no direct measurements for $Z_b^{(\prime)} \to \Upsilon_J(1D) \pi$. As indicated in the introduction, only the branching ratios of $\Upsilon(5S) \to \Upsilon_J(1D) \pi^+ \pi^-$ can be obtained from the experimental data. According to our analysis, we find that the cascade decays $\Upsilon(5S) \to Z_b^{(\prime)} \pi \to \Upsilon(1D) \pi^+ \pi^-$ play essential roles in understanding the large branching ratios of $\Upsilon(5S) \to \Upsilon(1D)\pi^+ \pi^-$. Together with the decay processes $\Upsilon(5S)\to Z_b^{(\prime)} \pi$, we can roughly estimate the branching ratios of $\Upsilon(5S) \to \Upsilon(1D) \pi^+ \pi^-$, which can be compared with the corresponding experimental data.

Here, we vary the parameter $\alpha$ from 0.5 to 1.0 for both $Z_b$ and $Z_b^\prime$. In the left panel of Fig.~\ref{Fig:zb}, we present the branching ratios of $Z_b\to \Upsilon_J(1D)\pi$ depending on the model parameter $\alpha$, which are,
\begin{eqnarray}
&&\mathcal{B}[Z_b\rightarrow \Upsilon_1 \pi]=(2.62-8.19)\times10^{-2},\nonumber\\
&&\mathcal{B}[Z_b\rightarrow \Upsilon_2 \pi]=(1.01-3)\times10^{-2},\nonumber\\
&&\mathcal{B}[Z_b\rightarrow \Upsilon_3 \pi]=(1.59-4.8)\times10^{-2},
\end{eqnarray}
respectively. It should be noticed that the partial widths of $Z_b \to \Upsilon_J(1D) \pi$ are proportional to the square of the coupling constants $g_{Z_bBB^\ast}$, while the coupling constants are estimated by the partial widths of $Z_b \to B^\ast \bar{B}$. Thus the uncertainties of the width of $Z_b$ will not affect the branching ratios of $Z_b\to \Upsilon_J(1D) \pi $.

As shown in left panel of Fig.~\ref{Fig:zb}, the $\alpha$ dependence of estimated branching ratios are very similar. Thus, one can estimate the ratios of these branching ratios, which are expected to be model independent. Here, we define $B_{ij}$ as the ratio of $\mathcal{B}(Z_b\to \Upsilon_i(1D)\pi)$ and $\mathcal{B}(Z_b\to \Upsilon_j(1D)\pi)$. The $\alpha$ dependence of these ratios are presented in the right panel of Fig.~\ref{Fig:zb}, which are,
\begin{eqnarray}
	B_{21} &=& 0.37 \sim 0.39 ,\nonumber\\
	B_{31} &=& 0.59 \sim 0.61,\nonumber\\
	B_{23} &=& 0.62 \sim 0.63,
\end{eqnarray}
respectively.

Similar to the case of $Z_b$, we can investigate the decay processes $Z^\prime_b\rightarrow \Upsilon_J(1D)\pi$. The $\alpha$ dependence of the branching ratios are presented in the left panel of Fig.~\ref{Fig:zbp}, and the present estimations indicate that the branching ratios are,
\begin{eqnarray}
&&\mathcal{B}[Z^\prime_b\rightarrow \Upsilon_1 \pi]=(3.38-10.67)\times10^{-2},\nonumber\\
&&\mathcal{B}[Z^\prime_b\rightarrow \Upsilon_2 \pi]=(1.56-4.64)\times10^{-2},\nonumber\\
&&\mathcal{B}[Z^\prime_b\rightarrow \Upsilon_3 \pi]=(2.72-8.22)\times10^{-2},
\label{Eq:brzbp}
\end{eqnarray}
respectively. The ratios of the branching ratios of $Z_b^\prime \to \Upsilon_J(1D) \pi$ are presented in the right panel of Fig. \ref{Fig:zbp}, where one can find that these ratios are very weakly dependent on the model parameter $\alpha$, which is similar to the case of $Z_b$. The ratios are estimated to be,
\begin{eqnarray}
	B_{21} &=& 0.43 \sim 0.46 ,\nonumber\\
	B_{31} &=& 0.77 \sim 0.80,\nonumber\\
	B_{23} &=& 0.56 \sim 0.57,
\end{eqnarray}
respectively.

From the measured data, one can conclude that the branching ratios of $\Upsilon(5S) \to \Upsilon_J(1D) \pi^+ \pi^-$ are about $1\times 10^{-3}$. Our analysis indicates that the dominant contributions come from the cascade decay processes $\Upsilon(5S) \to Z_b^{(\prime) \pm } \pi^\mp  \to \Upsilon_J(1D) \pi^+ \pi^-$. Since the widths of $Z_b$ and $Z_b^\prime$ are much smaller than their mass difference,  the interference between $Z_b$ and $Z_b^{\prime}$ in $\Upsilon(5S) \to \Upsilon_J(1D) \pi^+ \pi^-$ are negligible as indicated in Appendix~\ref{Sec:App-B}. Therefore, one can approximately estimate the branching ratios of $\Upsilon(5S) \to \Upsilon_J(1D) \pi^+ \pi^-$ by,
\begin{eqnarray}
&&	\mathcal{B}(\Upsilon(5S) \to \Upsilon_J(1D) \pi^+ \pi^-)\nonumber\\ &\simeq&  2 \Big(\mathcal{B}(\Upsilon(5S) \to Z_b^+ \pi^-) \times \mathcal{B}(Z_b^+\to \Upsilon_J(1D) \pi^+) \nonumber\\ &&+ \mathcal{B}(\Upsilon(5S) \to Z_b^{\prime +} \pi^-) \times \mathcal{B}(Z_b^{\prime +}\to \Upsilon_J(1D) \pi^+) \Big),
\end{eqnarray}
where the factor $2$ comes from the contributions of $Z_b^{(\prime)-}$. In Ref.\cite{Wu:2018xaa}, the production ratios of $Z^{(\prime)}_b$ from the $\Upsilon(5S)$ decays have been investigated and the production ratios were estimated to be of order $1\%$ for both $Z_b$ and $Z_b^\prime$. Taking $\mathcal{B}(\Upsilon(5S) \to Z_b^{(\prime)} \pi) =1.0\%$, one can conclude,
\begin{eqnarray}
	\mathcal{B}[\Upsilon(5S) \to \Upsilon_1(1D) \pi^+\pi^-)] &\sim & (1.2 - 3.7) \times 10^{-3},\nonumber\\
	\mathcal{B}[\Upsilon(5S) \to \Upsilon_2(1D) \pi^+\pi^-)] &\sim & (0.5 - 1.5) \times 10^{-3},\nonumber\\
	\mathcal{B}[\Upsilon(5S) \to \Upsilon_3(1D) \pi^+\pi^-)] &\sim & (0.8 - 2.6) \times 10^{-3},
\end{eqnarray}
which are consistent with the experimental measurement by the Belle collaboration in order of magnitude. Moreover, our estimates also indicate that the dominant component of the structure in the dipion missing mass spectrum should be $\Upsilon_1(1D)$.

\section{Summary}

The bottomonium $\Upsilon(5S)$ has become an intriguing source of investigating lower bottomonia and bottomnium-like states. Two interesting bottomonium-like states had been observed in the dipion transitions between $\Upsilon(5S)$ and lower $S$-wave and $P$-wave bottomonia. According to the measurements of the Belle Collaboration, we can obtained the branching ratios of $\Upsilon(5S) \to \Upsilon(1D) \pi^+ \pi^-$, which are the same order as those of $\Upsilon(5S)\to \Upsilon(nS)\pi^+ \pi^-$.

 Different from the case of $\Upsilon(5S)\to \Upsilon(nS)\pi^+ \pi^-$, our analysis indicates that the bottom meson loop contributions to $\Upsilon(5S) \to \Upsilon(1D) \pi^+ \pi^-$ should be strongly suppressed. Accordingly, the dominant source of the anomalous decay widths of $\Upsilon(5S) \to \Upsilon(1D) \pi^+ \pi^-$ should be $Z_b^{(\prime)}$. In the present work, we have estimated the decay processes of $Z_b^{(\prime)} \to \Upsilon(1D) \pi$, where the bottom meson loops play a role of bridging the initial $Z_b^{(\prime)}$ and $\Upsilon(1D) \pi$ in the final states.

 Our estimations indicate that the branching ratios of $Z_b^{(\prime )} \to \Upsilon(1D) \pi$ should be of order $10^{-2}$, and $\mathcal{B}(Z_b^{(\prime )} \to \Upsilon_1(1D) \pi)>\mathcal{B}(Z_b^{(\prime )} \to \Upsilon_3(1D) \pi)>\mathcal{B}(Z_b^{(\prime )} \to \Upsilon_2(1D) \pi)$. The branching ratios of $Z_b^{(\prime)} \to \Upsilon_J(1D) \pi$ are dependent on the model parameter $\alpha$, however, the ratios of these branching ratios are almost model independent, which could be an important test of the reliability of the present estimation. With the estimated branching ratios of $Z_b^{(\prime)} \to \Upsilon_J(1D) \pi$ and the assumptions that the branching ratios of $\Upsilon(5S) \to Z_b^{(\prime)} \pi $ are $1\%$, we have obtained the branching ratios for $\Upsilon(5S) \to \Upsilon_J(1D) \pi^+ \pi^-$, which are consistent with the Belle data in order of magnitude.

Before the end of this work, it is worth mentioning that in Refs.~\cite{Chen:2011pv, Chen:2012yr, Bugg:2011jr}, the structures corresponding to $Z_b^{(\prime)}$ were merely considered as some kinds of kinematical effects, which are not related to any poles in the $T-$matrix. Nevertheless, the experimental measurements indicate that the coupling between the structure $Z_b^{(\prime)}$ and the open bottom meson pair is much stronger and thus the hidden bottom final states can still couple to the structure $Z_b^{(\prime)}$ via a final state interaction as shown in Fig.~\ref{Fig:Feyn}. Then, the current conclusions remain true at least qualitatively in the kinematical effect scenario.

\section*{Acknowledgement}
The authors would like to thank Feng-Kun Guo and Qian Wang for useful discussions. This project is partly supported by the National Natural Science Foundation of China under Grant No. 11775050.

\appendix
\section{Amplitudes for $Z_b^{(\prime)} \to \Upsilon_2 \pi$ and $Z_b^{(\prime)} \to \Upsilon_3 \pi$}\label{Sec:App-A}
The decay amplitudes for the $Z^{(\prime)}_b \to \Upsilon_2 \pi$ corresponding to diagrams (f)-(i) in Fig.~\ref{Fig:Feyn} are
\begin{eqnarray}
\mathcal{M}_{f}&=&i^3 \int\frac{d^4 q}{(2\pi)^4}\Big[g_{Z_b}\epsilon^\rho_{Z_b}\Big]\Big[-i g_{B^\ast B\pi}ip_{3\sigma}\Big]\Big[ig_{\Upsilon_2 B^\ast B^\ast}\varepsilon_{\alpha\beta\mu\nu}\nonumber\\
&&\Big(g^\nu_\xi g_{\lambda\tau}(p^\beta_2-q^\beta)-g^\nu_\tau g_{\lambda\xi}(q^\beta-p^\beta_2)\Big)p^\mu_4 \epsilon^{\alpha\lambda}_{\Upsilon_2}\Big]\frac{1}{p^2_1-m^2_1}\nonumber\\
&&\frac{-g_\rho^{\tau}+p_{2\rho} p^{\tau}_2 /m^2_2}{p^2_2-m^2_2}\frac{-g^{\sigma\xi}+q^{\sigma}q^{\xi}/m^2_q}{q^2-m^2_q}\mathcal{F}^2(q^2,m_q^2)
\end{eqnarray}
\begin{eqnarray}
\mathcal{M}_{g}&=&i^3 \int\frac{d^4 q}{(2\pi)^4}\Big[g_{Z_b}\epsilon^\rho_{Z_b}\Big]\Big[\frac{1}{2}g_{B^\ast B^\ast \pi}\varepsilon_{\mu\nu\alpha\beta}ip^\nu_3 i(p^\alpha_1+q^\alpha)\Big]\nonumber\\
&&\Big[-ig_{\Upsilon_2 B^\ast B}\epsilon^{\tau\xi}_{\Upsilon_2}(-i)(p_{2\xi}-q_\xi)\Big]\frac{-g_\rho^{\mu}+p_{1\rho} p^{\mu}_1 /m^2_1}{p^2_1-m^2_1}\nonumber\\
&&\frac{1}{p^2_2-m^2_2}\frac{-g^{\beta}_\tau+q^{\beta}q_{\tau}/m^2_q}{q^2-m^2_q}\mathcal{F}^2(q^2,m_q^2)
\end{eqnarray}
\begin{eqnarray}
\mathcal{M}_{h}&=&i^3 \int\frac{d^4 q}{(2\pi)^4}\Big[ig_{Z^\prime_b}\varepsilon_{\mu\nu\alpha\beta}(-i)p^\mu \epsilon^\nu_{Z^\prime_b}\Big]\Big[ig_{B^\ast B \pi}ip_{3\rho}\Big]\nonumber\\
&&\Big[ig_{\Upsilon_2 B^\ast B}\epsilon^{\tau\xi}_{\Upsilon_2}(-i)(p_{2\xi}-q_\xi)\Big]\frac{-g^{\alpha\rho}+p^{\alpha}_1 p^{\rho}_1 /m^2_1}{p^2_1-m^2_1}\nonumber\\
&&\frac{-g^{\beta}_\tau+p^{\beta}_2 p_{2\tau} /m^2_2}{p^2_2-m^2_2}\frac{1}{q^2-m^2_q}\mathcal{F}^2(q^2,m_q^2)
\end{eqnarray}
\begin{eqnarray}
\mathcal{M}_{i}&=&i^3 \int\frac{d^4 q}{(2\pi)^4}\Big[ig_{Z^\prime_b}\varepsilon_{\mu\nu\alpha\beta}(-i)p^\mu \epsilon^\nu_{Z^\prime_b}\Big]\Big[\frac{1}{2}g_{B^\ast B^\ast \pi}\varepsilon_{\rho\tau\sigma\xi}ip^\tau_3 i\nonumber\\
&&(p^\sigma_1+q^\sigma)\Big]\Big[ig_{\Upsilon_2 B^\ast B^\ast}\varepsilon_{\theta\phi\kappa\lambda}\Big(g^\lambda_\eta g_{\delta\omega}(p^\phi_2-q^\phi)-g^\lambda_\omega g_{\delta\eta}(q^\phi\nonumber\\
&&-p^\phi_2)\Big)p^\kappa_4 \epsilon^{\theta\delta}_{\Upsilon_2}\Big]\frac{-g^{\alpha\rho}+p^{\alpha}_1 p^{\rho}_1 /m^2_1}{p^2_1-m^2_1}\frac{-g^{\beta\omega}+p^{\beta}_2 p^{\omega}_2 /m^2_2}{p^2_2-m^2_2}\nonumber\\
&&\frac{-g^{\xi\eta}+q^{\xi}q^{\eta}/m^2_q}{q^2-m^2_q}\mathcal{F}^2(q^2,m_q^2)
\end{eqnarray}

The decay amplitudes for the $Z^{(\prime)}_b \to \Upsilon_3 \pi$ corresponding to diagrams (j)-(k) in Fig.~\ref{Fig:Feyn} read
\begin{eqnarray}
\mathcal{M}_{j}&=&i^3 \int\frac{d^4 q}{(2\pi)^4}\Big[g_{Z_b}\epsilon^\rho_{Z_b}\Big]\Big[-i g_{B^\ast B\pi}ip_{3\sigma}\Big]\Big[g_{\Upsilon_3 B^\ast B^\ast}\epsilon^{\mu\nu\alpha}_{\Upsilon_3}(-i)\nonumber\\
&&(p_{2\mu}-q_\mu)(g_{\alpha\theta} g_{\nu\phi}+g_{\nu\theta} g_{\alpha\phi})]\frac{1}{p^2_1-m^2_1}\frac{-g_\rho^{\phi}+p_{2\rho} p^{\phi}_2 /m^2_2}{p^2_2-m^2_2}\nonumber\\
&&\frac{-g^{\sigma\theta}+q^{\sigma}q^{\theta}/m^2_q}{q^2-m^2_q}\mathcal{F}^2(q^2,m_q^2)
\end{eqnarray}
\begin{eqnarray}
\mathcal{M}_{k}&=&i^3 \int\frac{d^4 q}{(2\pi)^4}\Big[ig_{Z^\prime_b}\varepsilon_{\mu\nu\alpha\beta}(-i)p^\mu \epsilon^\nu_{Z^\prime_b}\Big]\Big[\frac{1}{2}g_{B^\ast B^\ast \pi}\varepsilon_{\rho\tau\sigma\xi}ip^\tau_3 i\nonumber\\
&&(p^\sigma_1+q^\sigma)\Big]\Big[g_{\Upsilon_3 B^\ast B^\ast}\epsilon^{\theta\phi\kappa}_{\Upsilon_3}(-i)(p_{2\theta}-q_\theta)(g_{\kappa\lambda} g_{\phi\eta}+g_{\phi\lambda} g_{\kappa\eta})\Big]\nonumber\\
&&\frac{-g^{\alpha\rho}+p^{\alpha}_1 p^{\rho}_1 /m^2_1}{p^2_1-m^2_1}\frac{-g^{\beta\eta}+p^{\beta}_2 p^{\eta}_2 /m^2_2}{p^2_2-m^2_2}\nonumber\\
&&\frac{-g^{\xi\lambda}+q^{\xi}q^{\lambda}/m^2_q}{q^2-m^2_q}\mathcal{F}^2(q^2,m_q^2)
\end{eqnarray}

\section{$\Upsilon_1 \pi^+$ invariant mass distribution of $\Upsilon(5S) \to \Upsilon_1 \pi^+ \pi^-$}\label{Sec:App-B}

\begin{figure}[htb]
  \centering
  \begin{tabular}{cc}
  	\includegraphics[width=3.5cm]{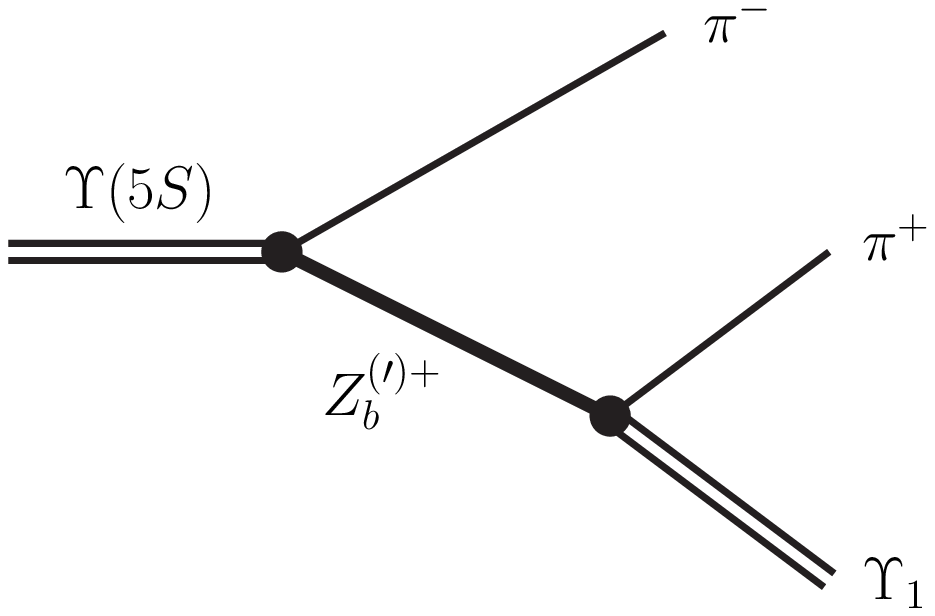}&
  	\includegraphics[width=3.5cm]{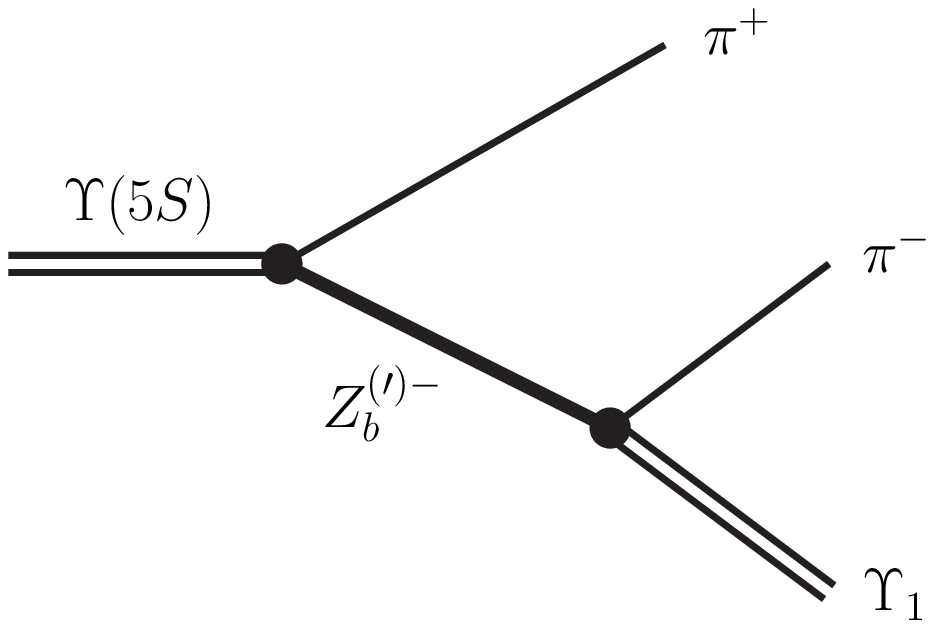}\\
  	$(a)$ & $(b)$
  \end{tabular}
   \caption{Cascade decay processes contributing to $\Upsilon(5S) \to \Upsilon_1 \pi^+\pi^-$.}\label{Fig:Feyn-Sim}
\end{figure}

The decay widths of $Z_b^{(\prime)}$ are much smaller than the mass splitting between these two states, thus the interference between $Z_b$ and $Z_b^\prime$ should be small and negligible. Here, we take $\Upsilon(5S) \to \Upsilon_1 \pi^+ \pi^-$ as an example to explicitly show the interference effect. In Fig.~\ref{Fig:Feyn-Sim}, we present the diagrams of  the cascade decay processes $\Upsilon(5S) \to Z_b^{(\prime) \pm } \pi^\mp \to \Upsilon_1 \pi^+ \pi^-$, which are considered as the primary sources of $\Upsilon(5S) \to \Upsilon_1 \pi^+ \pi^-$. For simplicity, we only consider the $S$-wave coupling in  both $\Upsilon(5S) Z_b^{(\prime)} \pi $ and $Z_b^{(\prime)} \Upsilon_1 \pi$ vertexes. Then, the amplitudes corresponding to the diagrams in Fig.~\ref{Fig:Feyn-Sim} for $Z_b$ read,
\begin{eqnarray}
\mathcal{A}_{Z_b}^{(a)} &=& \left[g_{\Upsilon(5S) Z_b \pi}\epsilon_{\Upsilon(5S)}^\mu \right]	\left[g_{Z_b\Upsilon_1 \pi}\epsilon_{\Upsilon_1}^\nu \right] \nonumber\\
&\times&  \frac{-g_{\mu \nu}+ (p_{2\mu}+p_{3\mu})(p_{2\nu}+p_{3\nu}) /m_{Z_b}^2}{(p_2+p_3)^2-m_{Z_b}^2+im_{Z_b} \Gamma_{Z_b}}\nonumber\\
\mathcal{A}_{Z_b}^{(b)} &=& \left[g_{\Upsilon(5S) Z_b \pi}\epsilon_{\Upsilon(5S)}^\mu \right]	\left[g_{Z_b\Upsilon_1 \pi}\epsilon_{\Upsilon_1}^\nu \right] \nonumber\\
&\times&  \frac{-g_{\mu \nu}+ (p_{1\mu}+p_{3\mu})(p_{1\nu}+p_{3\nu}) /m_{Z_b}^2}{(p_1+p_3)^2-m_{Z_b}^2+im_{Z_b} \Gamma_{Z_b}}.
\end{eqnarray}
Similarly, one obtains the amplitudes corresponding to the contribution from $Z_b^{\prime}$. Then the contribution from $Z_b^{(\prime)}$ is
\begin{eqnarray}
A_{Z_b^{(\prime)}}=	A_{Z_b^{(\prime)}}^{(a)}+A_{Z_b^{(\prime)}}^{(b)}.
\end{eqnarray}
Here, the coupling constants $g_{\Upsilon(5S) Z_b^{(\prime)} \pi}$ can be estimated from the assumption that the branching ratios of $\Upsilon(5S) \to Z_b^{(\prime)} \pi $ are $1\%$, while $g_{Z_b^{(\prime)} \Upsilon_1 \pi}$ are estimated from the partial widths given in the present work with $\alpha= 0.5$. The partial decay width reads
\begin{eqnarray}
d\Gamma= \frac{1}{3} \frac{1}{32 m_{0}^3} \overline{|A_{\mathrm{tot}}|^2} dm_{12}^2 dm_{23}^2
\end{eqnarray}
where $A_{\mathrm{Tot}}=A_{Z_b} +A_{Z_b^\prime}$. Then, we have
\begin{eqnarray}
	|A_{\mathrm{Tot}}|^2=|A_{Z_b} +A_{Z_b^\prime}|^2 ,\label{Eq:Interfer}
\end{eqnarray}
where the interference between $Z_b$ and $Z_b^\prime$ has been included. Neglecting the interference between $Z_b$ and $Z_b^\prime$, one can approximately obtain the square of the amplitude as
\begin{eqnarray}
	|A_{\mathrm{Tot}}|^2 \simeq |A_{Z_b}|^2 +|A_{Z_b^\prime}|^2 .\label{Eq:Interfer}
\end{eqnarray}

\begin{figure}[htb]
  \centering
 \includegraphics[width=8cm]{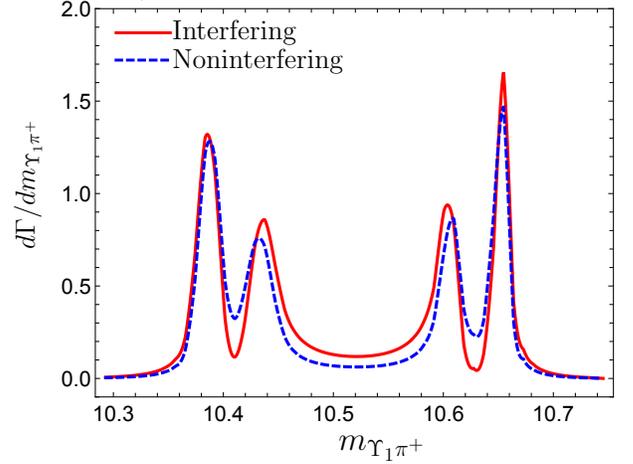}
  \caption{The $\Upsilon_1 \pi^+$ invariant mass distribution of $\Upsilon(5S)\rightarrow\Upsilon_1(1D)\pi^+ \pi^-$. The solid and dashed curves are the distributions with and without the interference between $Z_b$ and $Z_b^\prime$.}\label{Fig:Compare}
\end{figure}

In Fig.~\ref{Fig:Compare}, we present the $\Upsilon_1 \pi^+$ distributions obtained with and without the interference between $Z_b$ and $Z_b^\prime$, where one can find the curves with and without the interference are very similar. Thus, one can consider the case without the interference as a good approximation in the estimations, which is similar to the case of $\Upsilon(5S) \to h_b \pi^+ \pi^-$.

\end{document}